\documentclass[showpacs]{revtex4}
\usepackage{epsfig}
\usepackage{graphicx}
\usepackage{dcolumn}
\usepackage{amsmath}
\usepackage{latexsym}
\usepackage{amsfonts}
\usepackage{amssymb}
\usepackage{slashed}
\usepackage{braket}
\usepackage{physics}
\begin{document}

%%%%%%%%%%%%%%%%%%%%
\title{Fuzzy Classical Dynamics as a Paradigm for Emerging Lorentz Geometries}  
%\date{\today}
\author{F.G. Scholtz$^{a,b}$, P. Nandi$^c$, S.K. Pal$^c$ and B. Chakraborty$^c$ }
\affiliation{$^a$National Institute for Theoretical Physics (NITheP), 
Stellenbosch 7602, South Africa\\
$^b$Institute of Theoretical Physics, 
Stellenbosch University, Stellenbosch 7602, South Africa\\
$^c$S.N.Bose National Centre For Basic Sciences, JD Block, Salt Lake, Kolkata-700106, India}

%%%%%%%%%%%%%%%%
\begin{abstract}
\noindent 
 We show that the classical equations of motion for a particle on three dimensional fuzzy space and on the fuzzy sphere are underpinned by a natural Lorentz geometry.  From this geometric perspective, the equations of motion generally correspond to forced geodesic motion, but for an appropriate choice of noncommutative dynamics, the force is purely noncommutative in origin and the underpinning Lorentz geometry some standard space-time with, in general,  non-commutatuve corrections to the metric.  For these choices of the noncommutative dynamics the commutative limit therefore corresponds to geodesic motion on this standard space-time.  We identify these Lorentz geometries to be a Minkowski metric on $\mathbb{R}^4$ and $\mathbb{R} \times S ^2$ in the cases of a free particle on three dimensional fuzzy space ($\mathbb{R}^3_\star$) and the fuzzy sphere ($S^2_\star$), respectively. We also demonstrate the equivalence of the on-shell dynamics of $S^2_\star$ and a relativistic charged particle on the commutative sphere coupled to the background magnetic field of a Dirac monopole. 
\end{abstract}

\maketitle

%%%%%%%%%%%%%%%%%%%%%%%%%%%%%%%%%%%

%%%%%%%%%%%%%%%%%%%%%%%%%%%%%%%%%%%%%%%%%%%%%%%%%%%%%%%%%%%%

\section{Introduction}
\label{intro}

The structure of space-time at short length scales and the emergence of space-time as we perceive it at long length scales are probably the most challenging problems facing modern physics \cite{seib}.  These issues are also at the core of the struggle to combine gravity and quantum mechanics into a unified theory and probably also link closely with the observational challenges of dark matter and energy.

One scenario for space-time at short length scales is that of noncommutative space-time, which has received considerable attention in the past few decades.  This was originally proposed by Snyder \cite{Snyder} in an attempt to avoid the ultra-violet infinities of field theories.  The discovery of renormalization pushed these ideas to the background until more recently when they resurfaced in the search for a consistent theory of quantum gravity.  The compelling arguments of Doplicher et al \cite{dop} highlighted the need for a revised notion of space-time at short length scales and gave strong arguments in favour of a noncommutative geometry.  Shortly thereafter it was also noted that noncommutative coordinates occurred quite naturally in certain string theories \cite{wit}, generally perceived to be the best candidate for a theory of quantum gravity.  This sparked renewed interest in noncommutative space-time and the formulation of quantum mechanics \cite{scholtz} and quantum field theories on such spaces \cite{doug}.  

 In a recent paper \cite{scholtz1} the equations of motion for a particle moving in three dimensional fuzzy space were derived and the most striking features of these equations were the emergence of a limiting speed and energy.  In form these equations are also very reminiscent to relativistic dynamics and it is natural to ask to what extent these equations are in fact underpinned by some Lorentz geometry.  The purpose of the present paper is to explore this possibility.  We explicitly demonstrate that there is in fact a natural Lorentz geometry underpinning these equations that can simply be identified on the level of the equations of motion.  We then show that these equations admit the natural interpretation of geodesic equations in the presence of a force.  We continue to show that for an appropriate choice of noncommutative dynamics, the force is purely noncommutative and vanishes in the commutative limit, where the equations of motion are geodesic.   We demonstrate the above in two cases namely, three dimensional fuzzy space ($\mathbb{R}^3_\star$ ) and the fuzzy sphere ($S^2_\star$).  We also show that the on-shell dynamics on the fuzzy sphere is equivalent to the on-shell dynamics of a charged particle on the commutative sphere coupled to the background magnetic field of a Dirac monopole.  A similar construction is currently lacking for three dimensional fuzzy space and it is unclear whether it even exists.
  
 This paper is organised as follows: Section \ref{NCQM} collects the main results regarding the formulation of quantum mechanics on three dimensional fuzzy space and the fuzzy sphere.  A detailed exposition of the formulation of quantum mechanics on the Moyal plane can be found in \cite{scholtz} and the generalisation to three dimensional fuzzy space in \cite{scholtz2}.  In \cite{scholtz1} a review of this formulation as well as a detailed derivation of the path integral action for a particle moving in three dimensional fuzzy space has been given and we refer the reader for the details to \cite{scholtz1}.  Sections \ref{fuzzyspace} and \ref{fuzzysphere} respectively discuss the geometric interpretation of the classical equations of motions on three dimensional fuzzy space and the fuzzy sphere.  In contrast to the path integral action of section \ref{fuzzysphere} for $S^2_\star$ that contains auxiliary variables, section \ref{action} constructs an action involving only the dynamic degrees of freedom and yields the same equations of motion for the fuzzy sphere. We also elaborate on the relation between the constants of motion of the two theories.  Section \ref{Conclusions} summarises the main findings and highlights open issues.

\section{Quantum mechanics on three dimensional fuzzy space ($\mathbb{R}^3_\star$ ) and the fuzzy sphere ($S^2_\star$)}
\label{NCQM}

The starting point is the fuzzy sphere commutation relations
\begin{equation}
[\hat{x}_i,\hat{x}_j]=2i\lambda\varepsilon_{ijk}\hat{x}_k,
\end{equation}
where $\lambda$ has the units of a length and $\varepsilon_{ijk}$ is the standard completely anti-symmetric tensor.  This coordinate algebra is then realised using the standard Schwinger realisation of $SU(2)$.  
\begin{equation}
\label{schw}
\hat{x}_i=\lambda a^\dagger_\alpha \sigma^{(i)}_{\alpha\beta} a_\beta.
\end{equation}
Here a summation over repeated indices is implied, $\alpha,\beta=1,2$, $\sigma^{(i)}_{\alpha\beta}$, $i=1,2,3$ are the Pauli spin matrices and $a_\alpha^\dagger$, $a_\alpha$ are standard boson creation and annihilation operators.  The radius operator is 
\begin{equation}
\label{radsq}
\hat{r}^2=\hat{x}_i\hat{x}_i =\lambda^2 \hat{n}(\hat{n}+2),
\end{equation}
with $\hat{n}=a_\alpha^\dagger a_\alpha$ (summation implied) the boson number operator.  Note that the radius operator is also the Casimir of $SU(2)$ and commutes with the coordinates.    As a measure of the radius we use 
\begin{equation}
\label{radiusc}
\hat{r}=\lambda (\hat{n}+1),
\end{equation}
which is to leading order the square root of (\ref{radsq}). We denote the two mode boson Fock space introduced above by ${\cal H}_c$ and denote its elements by $|n_1,n_2\rangle$ with $n_1,\,n_2$ the eigenvalues of the number operators $a^\dagger_1a_1,\, a^\dagger_2 a_2$.

The next step is to introduce the Hilbert space of Hilbert-Schmidt operators on ${\cal H}_c$,  i.e. the space of all operators generated by the creation and annihilation operators and with finite Hilbert-Schmidt norm.  This space is denoted by ${\cal H}^0_q$, states in this space by $|\psi)$ with $\psi$ a Hilbert-Schmidt operator on ${\cal H}_c$ and the inner product is given by
\begin{equation}
\label{innp1}
	(\tilde{\psi}|\tilde{\phi})={\rm tr}_c(\tilde{\psi}^\dagger\tilde{\phi}).
\end{equation}

Next one introduces operators on ${\cal H}^0_q$ (often referred to as super operators in the literature).  We reserve capitals to distinguish them from their counterparts acting on ${\cal H}_c$.  In doing this one notes that since the elements of ${\cal H}^0_q$ are operators, one must distinguish between left and right multiplication.  It is therefore convenient to introduce the following notation 
\begin{eqnarray}
&&A_{\alpha L}^\dagger|\psi)=|a_\alpha^\dagger\psi),\quad A_{\alpha L}|\psi)=|a_\alpha\psi),\nonumber\\
&&A_{\alpha R}^\dagger|\psi)=|\psi a_\alpha^\dagger),\quad A_{\alpha R}|\psi)=|\psi a_\alpha).
\end{eqnarray}
It is easily checked that these operators are hermitian conjugates with respect to the inner product (\ref{innp1}).  Also note that 
\begin{equation}
[A_{\alpha L},A_{\beta L}^\dagger]=\delta_{\alpha\beta},\quad [A_{\alpha R},A_{\beta R}^\dagger]=-\delta_{\alpha\beta}
\end{equation}
and all other commutators vanish.  All observables acting on ${\cal H}^0_q$ can now be constructed from these operators.  

Not all the states in ${\cal H}^0_q$ are, however, physical.  The physical states are those generated by the coordinates $\hat{x}_i$ only and are characterised by the condition 
\begin{equation}
\hat\Gamma|\psi)\equiv\left(A_{\alpha L}^\dagger A_{\alpha L}-A_{\alpha R} A_{\alpha R}^\dagger\right)|\psi)=|[a^\dagger_\alpha a_\alpha,\psi])=0.
\end{equation}
We denote this physical subspace by ${\cal H}_q$.  All physical observables must respect this constraint (leave ${\cal H}_q$ invariant) and therefore commute with $\hat{\Gamma}$.  

It is convenient to introduce the observables
\begin{eqnarray}
\hat{X}_{i L}=\lambda A^\dagger_{\alpha L}\sigma^{(i)}_{\alpha\beta} A_{\beta L},\nonumber\\
\hat{X}_{i R}=\lambda A_{\alpha R}{\sigma^\dagger}^{(i)}_{\alpha\beta}  A_{\beta R}^\dagger.
\end{eqnarray}
Note that left and right observables commute.  In terms of these, the observables of interest to us are 
\begin{eqnarray}
\label{obs}
\hat{X}_{i }&=&\hat{X}_{i L}\quad (\textrm{coordinates})\nonumber\\
\hat{R}&=&\lambda \left(A^\dagger_{\alpha L} A_{\alpha L}+1\right)\equiv\lambda(\hat{N}_L+1)\quad (\textrm{radius})\nonumber\\
\hat{L}_i&=&\frac{\hbar}{2\lambda}\left(\hat{X}_{i L}-\hat{X}_{i R}\right)\quad (\textrm{angular momentum}).
\end{eqnarray}

The Hamiltonian is taken to be \cite{scholtz1}
\begin{equation}
\label{ham3}
\hat{H}=\frac{\hbar^2}{2m\lambda^2}\left(A_{\alpha L}^\dagger f_1(\hat{R})A_{\alpha L}-A_{\alpha L}^\dagger f_2(\hat{R})A_{\alpha R}-A_{\alpha R}^\dagger f_2(\hat{R}) A_{\alpha L} +A_{\alpha R} ^\dagger f_3(\hat{R}) A_{\alpha R}\right)+V(\hat{R})\equiv\frac{\hbar^2}{2m}\hat{\Delta}+V(\hat{R})
\end{equation}
Here $f_i(\hat{R})$ are arbitrary real dimensionless functions, which generalises the Laplacian, $\hat{\Delta}$, in \cite{scholtz1}.  It is obvious that all these observables are hermitian and commute with $\hat{\Gamma}$. It can also be readily verified that the angular momentum operators commute with the Hamiltonian.  

The Laplacian of \cite{scholtz1} corresponds to the choice of functions
\begin{eqnarray}
\label{flat}
f_1(\hat R)&&=\frac{1}{\hat{N}_L+2},\nonumber\\
f_2(\hat R)&&=\frac{1}{\sqrt{(\hat{N}_L+1)(\hat{N}_L+2)}},\nonumber\\
f_3(\hat R)&&=\frac{1}{\hat{N}_L+1}.
\end{eqnarray}
After applying the similarity transformation $S^{-1}\hat{\Delta}S$ with $S=\sqrt{\hat{N}_L+1}$, this Laplacian can be recast into the form \cite{scholtz1,scholtz2}
\begin{equation}
 \hat{\Delta}|\psi)=|\frac{1}{\lambda\hat{r}}[\hat{a}^\dagger_\alpha,[\hat{a}_\alpha,\psi]])=|\frac{1}{\lambda^2\left(\hat{n}+1\right)}[\hat{a}^\dagger_\alpha,[\hat{a}_\alpha,\psi]]).
\end{equation}
The spectrum of the corresponding free particle Hamiltonian ($V(\hat{R})=0$) can be computed quite easily (see \cite{scholtz3} for a simple derivation) to find 
\begin{equation}
\label{spectrum}
E(\vec{k})=\frac{2\hbar^2}{m\lambda^2}\sin^2\left(\frac{|\vec{k}|\lambda}{2}\right)
\end{equation}
where $|\vec{k}|<\frac{\pi}{\lambda}$. Note the bound on the energy $E(\vec{k})\leq\frac{2\hbar^2}{m\lambda^2}$.  At low energies or in the commutative limit, this yields the non-relativistic dispersion relation, which motivates the choice of Hamiltonian for a free particle on three dimensional fuzzy space to be 
\begin{equation}
\label{free3d}
\hat{H}_0=\frac{\hbar^2}{2m\lambda^2}\left[A_{\alpha L}^\dagger \frac{1}{\hat{N}_L+2}  A_{\alpha L}-A_{\alpha L}^\dagger \frac{1}{\sqrt{(\hat{N}_L+1)(\hat{N}_L+2)}} A_{\alpha R}-A_{\alpha R}^\dagger \frac{1}{\sqrt{(\hat{N}_L+1)(\hat{N}_L+2)}} A_{\alpha L} +A_{\alpha R} ^\dagger  \frac{1}{\hat{N}_L+1} A_{\alpha R}\right].
\end{equation}

Next we consider the quantum mechanics of a particle on the fuzzy sphere, i.e. a rotor.  The construction of this noncommutative quantum system follows the same route as outlined above for three dimensional fuzzy space, the only difference being that now only one fuzzy sphere is considered.  Invoking the Schwinger representation (\ref{schw}), the configuration space is simply
\begin{equation}
{\cal H}_c = \textrm{span}\{ |j,m\rangle\}_{m=-j}^j,
\label{spherc}
\end{equation}
where $j=\frac{n_1+n_2}{2}=\frac{n}{2}$ and $m=\frac{n_1-n_2}{2}$.  Note that $j$ is fixed and determines the radius of the fuzzy sphere to be $r=\lambda(2j+1)=\lambda(n+1)$.  The quantum Hilbert space is now the space of all Hilbert-Schmidt operators on ${\cal H}_c$ and since, by construction, they cannot change the total number of bosons, they are automatically restricted to the operators generated by the coordinates and no further constraints need to be imposed:
\begin{equation}
{\cal H}_q=\textrm{span}\{|j,m_1\rangle\langle j,m_2|\}_{m_1,m_2=-j}^j.
\end{equation}
We denote the elements of $H_q$ by $|j,m_1,m_2)$ and the inner product is still given by (\ref{innp1}).  

All observables are constructed as in (\ref{obs}), but the Hamiltonian is now taken to be the free particle Hamiltonian
\begin{equation}
\label{freesp}
\hat{H}_0= \frac{\hat{L}^2}{2mr^2},
\end{equation}
which can also be rewritten, by using (\ref{obs}), as
\begin{eqnarray}
\label{hams}
\hat{H}_0&=& \frac{\hbar^2}{8m\lambda^2  r^2} \bigg[(\vec{X}^L)^2 +(\vec{X}^R)^2 -2 \vec{X}^L . \vec{X}^R\bigg]\nonumber\\
&=&\frac{\hbar^2}{4m\lambda^2}- \frac{\hbar^2}{4m\lambda^2  r^2} \vec{X}^L \cdot \vec{X}^R.
\end{eqnarray}
 Here we have used the fact that both $(\vec{X}^L)^2$ and $(\vec{X}^R)^2$ are Casimir operators and take the same constant value $\lambda^2 n(n+2)\approx r^2$.  For convenience, we shall suppress this constant in the action as it does not affect the equations of motion, but it is an important contribution to the energy. 

\section{Classical dynamics on three dimensional fuzzy space ($\mathbb{R}^3_\star$ )}
\label{fuzzyspace}

We introduce the standard minimum uncertainty states on ${\cal H}_c$ as Glauber coherent states, which form an overcomplete basis
\begin{eqnarray}
&&|z_\alpha\rangle=e^{-\bar{z}_\alpha z_\alpha/2}e^{z_\alpha a^\dagger_\alpha}|0\rangle,\nonumber\\
&&\int \frac{d\bar{z}_\alpha dz_\alpha}{\pi^2} |z_\alpha\rangle\langle z_\alpha|=\bf{1}_c.
\end{eqnarray}
The dimensionful physical coordinates are then identified as
\begin{equation}
\label{coord}
x_i=\langle z_\alpha |\hat{x}_i| z_\alpha\rangle=\lambda \bar{z}_\alpha \sigma^{(i)}_{\alpha\beta} z_\beta.
\end{equation}

Correspondingly, we introduce coherent states on ${\cal H}^0_q$ as
\begin{equation}
|z_\alpha, w_\alpha)=|z_\alpha\rangle\langle w_\alpha|.
\end{equation}
They are overcomplete and 
\begin{equation}
\int \frac{d\bar{z}_\alpha dz_\alpha d\bar{w}_\alpha dw_\alpha}{\pi^4} |z_\alpha,w_\alpha)(z_\alpha, w_\alpha|={\bf 1}^0_q.
\end{equation}

The derivation of the path integral action proceeds as detailed in \cite{scholtz1}.  It is convenient to work in dimensionless units by introducing the time scale, $t_0$, energy scale, $e_0$, dimensionless time, $T$, dimensionless coordinates, $X_i$, and dimensionless energy, $E$, as follow:
\begin{equation}
\label{dim}
t_0=\frac{m\lambda^2}{\hbar},\quad e_0=\frac{\hbar}{t_0},\quad T=\frac{t}{t_0},\quad X_i=\frac{x_i}{\lambda},\quad E=\frac{e}{e_0}.
\end{equation}
In what follows we reserve capitals for dimensionless quantities.  Note that these should not be confused with operators on ${\cal H}^0_q$ that are distinguished by a hat.
The dimensionless path integral action is found to be \cite{scholtz1}
\begin{equation}
\label{action3d}
S=\int_{T_i}^{T_f}dT \left[\frac{i}{2}\left(\bar{z}_\alpha\dot{z}_\alpha-\dot{\bar{z}}_\alpha z_\alpha+\dot{\bar{w}}_\alpha w_\alpha-\bar{w}_\alpha\dot{w}_\alpha\right)-H(z_\alpha,\bar{z}_\alpha,w_\alpha,\bar{w}_\alpha)\right],
\end{equation}
where the over-dot denotes derivation with respect to $T$ and
\begin{equation}
\label{action3dh}
H(z,\bar{z},w,\bar{w})=\left(f_1(R)\bar{z}_\alpha z_\alpha-f_2(R)\left(\bar{z}_\alpha w_\alpha+z_\alpha \bar{w}_\alpha\right)+f_3(R)\bar{w}_\alpha w_\alpha\right)+W(R),
\end{equation}
\begin{eqnarray}
\label{action3da}
f_i(R)&&=\frac{1}{2}\langle z_\alpha|f_i(\hat{R}|z_\alpha\rangle,\nonumber\\
W(R)&&=\frac{1}{e_0}\langle z_\alpha|V(\hat{R})|z_\alpha\rangle+2f_3(R)\equiv \frac{V(R)}{e_0}+2f_3(R).
\end{eqnarray}
There are five dimensionless conserved quantities, four related to a $U(2)$ symmetry and the fifth a conserved energy related to time translation invariance. These are easily found to be
\begin{eqnarray}
\label{con3d}
\Gamma&=&\bar{z}_\alpha z_\alpha-\bar{w}_\alpha w_\alpha,\nonumber\\
L_i&= &\bar{z}_\alpha \sigma^{(i)}_{\alpha\beta} z_\beta-\bar{w}_\alpha \sigma^{(i)}_{\alpha\beta} w_\beta,\nonumber\\
E&=&\tilde{H}(z,\bar{z},w,\bar{w}).
\end{eqnarray}

The condition of physicality of the initial and final states requires $\Gamma=0$.  Taking this into account, the equation of motion for the dimensionless physical coordinates $X_i$ can now be extracted as detailed in \cite{scholtz1}.  The result is 
\begin{equation}
\label{eqmf}
\ddot{\vec{X}}_\pm=a_\pm(R,V) \vec{X}+b_\pm(R,V)\left(\vec{X}\times\dot{\vec{X}}\right)+c_\pm(R)\left(\left(\vec{X}\times\dot{\vec{X}}\right)\times\dot{\vec{X}}\right).
\end{equation}
where $R^2=\vec{X}\cdot\vec{X}$ and 
\begin{eqnarray}
\label{eqmfa}
a_\pm(R,V)&=&4R^2f_2(R)^2g_1(R)\pm\frac{1}{R}{\frac{d g_2(R)}{dR}}\sqrt{4R^2f_2(R)^2-\dot{\vec{X}}\cdot\dot{\vec{X}}},\nonumber\\
b_\pm(R,V)&=&\frac{1}{R}\frac{d g_2(R)}{dR}\pm g_1(R)\sqrt{4R^2f_2(R)^2-\dot{\vec{X}}\cdot\dot{\vec{X}}},\nonumber\\
c_\pm(R)&=&g_1(R).
\end{eqnarray}
Here
\begin{eqnarray}
\label{eqmfaa}
g_1(R)&=&\frac{1}{R^2}+\frac{1}{f_2(R)R}\frac{d f_2 (R)}{dR},\nonumber\\
g_2(R)&=&R\left(f_1(R)+f_3(R)\right)+W(R).
\end{eqnarray}

The dimensionless conserved quantities can also be computed, but now there are only four as the constraint $\Gamma=0$ is satisfied by construction.  They are
\begin{eqnarray}
\label{const}
\vec{L}_\pm&=&\frac{1}{4f_2(R)^2R^2}\left[\sqrt{4R^2f_2(R)^2-\dot{\vec{X}}\cdot\dot{\vec{X}}}\left(\vec{X}\times\dot{\vec{X}}\right)\pm \left(\vec{X}\times\dot{\vec{X}}\right)\times\dot{\vec{X}}\right],\label{consta}\\
E_\pm&=&g_2(R)\pm \sqrt{4R^2f_2(R)^2-\dot{\vec{X}}\cdot\dot{\vec{X}}}.\label{constb}
\end{eqnarray}
Note that there are two branches denoted $\pm$.  Only the minus branch reduces to the Newtonian equations of motion in the commutative limit and we focus on this branch from here on.  

Before continuing, it is necessary to make the $\lambda$ dependence in these equations explicit in order to study the commutative limit.  Restoring dimensions for the minus branch using (\ref{dim}) we have 
\begin{equation}
\label{eqmfd}
\frac{d^2\vec{x}}{dt^2}=\tilde{a}(r,v) \vec{x}+\tilde{b}(r,v)\left(\vec{x}\times\frac{d\vec{x}}{dt}\right)+\tilde{c}(r)\left(\left(\vec{x}\times\frac{d\vec{x}}{dt}\right)\times\frac{d\vec{x}}{dt}\right).
\end{equation}
Here the dimensionful coefficients are given by
\begin{eqnarray}
\label{eqmfdcon}
\tilde{a}(r,v,\lambda)&=&v_0^2\left[\frac{4r^2f_2^2(r,\lambda)g_1(r,\lambda)}{\lambda^4}- \frac{1}{r}\frac{d g_2(r,\lambda)}{dr}\sqrt{\frac{4r^2f_2^2(r,\lambda)}{\lambda^2}-\frac{1}{v_0^2}\left(\frac{d\vec{x}}{dt}\right)^2}\,\right]\nonumber\\
\tilde{b}(r,v,\lambda)&=&v_0\left[\frac{1}{r}\frac{d g_2(r,\lambda)}{dr}- \frac{g_1(r,\lambda)}{\lambda^2}\sqrt{\frac{4r^2f_2^2(r,\lambda)}{\lambda^2}-\frac{1}{v_0^2}\left(\frac{d\vec{x}}{dt}\right)^2}\,\right],\nonumber\\
\tilde{c}(r,\lambda)&=&\frac{g_1(r,\lambda)}{\lambda^2},
\end{eqnarray}
where we have introduced the limiting speed 
\begin{equation}\label{speedlimit}
v_0=\frac{\hbar}{m\lambda},
\end{equation}
and from (\ref{action3da}) and (\ref{eqmfaa})
\begin{eqnarray}
\label{eqmfda}
g_1(r,\lambda)&=&\lambda^2\left(\frac{1}{r^2}+\frac{1}{f_2(r,\lambda)r}\frac{d f_2 (r,\lambda)}{dr}\right),\nonumber\\
g_2(r,\lambda)&=&\frac{r}{\lambda}\left(f_1(r,\lambda)+f_3(r,\lambda)\right)+\frac{V(r,\lambda)}{mv_0^2}+2f_3(r,\lambda).
\end{eqnarray}

Similarly, the dimensionful energy $e=e_0E$ and angular momentum $\vec{\ell}=\hbar\vec{L}$ can be written as
\begin{eqnarray}
\label{constd}
\vec{\ell}&=&\frac{m\lambda^2}{4f_2^2(r,\lambda)r^2}\left[\sqrt{\frac{4r^2f_2^2(r,\lambda)}{\lambda^2}-\frac{1}{v_0^2}\left(\frac{d\vec{x}}{dt}\right)^2}\left(\vec{x}\times\frac{d\vec{x}}{dt}\right)-\frac{1}{v_0}\left(\left(\vec{x}\times\frac{d\vec{x}}{dt}\right)\times\frac{d\vec{x}}{dt}\right)\right],\label{constad}\\
e&=&mv_0^2\left[g_2(r,\lambda)-\sqrt{\frac{4r^2f_2^2(r,\lambda)}{\lambda^2}-\frac{1}{v_0^2}\left(\frac{d\vec{x}}{dt}\right)^2}\,\right].\label{constbd}
\end{eqnarray}

It is important to realise that there is no unique way of taking the classical ($\hbar\rightarrow 0$) and commutative ($\lambda\rightarrow 0$) limits and that the result depends on how it is done.  For example, taking the $\hbar\rightarrow 0$ limit before the $\lambda\rightarrow 0$ limit, leads to a nonsensical theory with $v_0=0$.  This suggests that one cannot give sensible meaning to a noncommutative classical theory.  This was also already observed in \cite{scholtz4} in the context of the nonlocal action of a particle on the Moyal plane.  Equations (\ref{eqmfd}), (\ref{eqmfdcon}), (\ref{eqmfda}) and (\ref{constd}),  however, suggest that the most natural way of taking this limit is such that $\frac{\hbar}{m\lambda}=v_0$ is fixed (note that if we want $v_0$ to be independent of $m$, this limit must be taken in an appropriate mass dependent way).  For a smooth limit one must then also require $f_i(r,\lambda)=\lambda \tilde{f}_i(r)+O(\lambda^2)$, which automatically implies  
\begin{eqnarray}
\label{exp}
g_1(r,\lambda)=\lambda^2\left(\frac{1}{r^2}+\frac{1}{r\tilde{f}_2(r)}\frac{d\tilde{f}_2(r)}{dr}\right)\equiv \lambda^2\tilde{g}_1(r)+O(\lambda^3),\nonumber\\
g_2(r,\lambda)=r\left(\tilde{f}_1(r)+\tilde{f}_3(r)\right)+\frac{\tilde{V}(r)}{mv_0^2}+O(\lambda)\equiv\tilde{g}_2(r)+O(\lambda),
\end{eqnarray}
where we have also introduced $V(r,\lambda)=\tilde{V}(r)+O(\lambda)$.  Under these conditions the $\hbar\rightarrow 0$ and $\lambda\rightarrow 0$ limits, such that $\frac{\hbar}{m\lambda}=v_0$ is fixed, can safely be taken with the result that $f_i(r,\lambda)$, $g_i(r,\lambda)$ simply get replaced by $\tilde{f}_i(r)$ and $\tilde{g}_i(r)$, respectively.  

The limit discussed above is clearly not the Newtonian limit, but rather leads one to a theory reminiscent of relativistic dynamics with a limiting speed.  As quantum mechanics on $\mathbb{R}^3_\star$ with Laplacian as in (\ref{flat}) reduces to standard quantum mechanics in the $\lambda\rightarrow 0$ limit \cite{scholtz2,scholtz3}, one expects the Newtonian limit to result from a $\lambda\rightarrow 0$ limit followed by a classical $\hbar\rightarrow 0$ limit or, alternatively, a $v_0\rightarrow\infty$ limit.  This is indeed the case. For the choice of functions $f_i(\hat{R})$ in (\ref{flat}) the functions $f_i(R)$ in (\ref{action3da}) can be computed as described in \cite{scholtz1} to find $\tilde{f}_i(r)=\frac{1}{2r}$, $\tilde{g}_1(r)=0$, $\tilde{g}_2(r)=1+\frac{\tilde{V}(r)}{mv_0^2}$.  Substituting this in (\ref{eqmfd}) and taking the  $\lambda\rightarrow 0$ limit followed by the $v_0\rightarrow\infty$ limit indeed yields
\begin{equation}
\frac{d^2\vec{x}}{dt^2}=-\frac{1}{mr}\frac{d\tilde{V}(r)}{dr}\vec{x},
\end{equation}
which is Newton's equation for a central potential.  For this reason we shall refer to the $v_0\rightarrow\infty$ limit as the commutative limit in what follows.  Note that this limit is reminiscent of a nonrelativistic limit, but we prefer not to refer to it as such to avoid possible confusion.

The observations above and the structure of the equations  (\ref{eqmfd}), (\ref{eqmfdcon}), (\ref{eqmfda}) and (\ref{constd}) naturally lead one to the question whether these equation may in fact admit a natural interpretation as covariant geodesic equations, possibly in the presence of a 4-force.   Inspecting these equations, one is naturally led to introduce, after restoration of dimensions, the dimensionful 4-vector $x^\mu=(v_0t,\vec{x})$ ($\mu=0,1,2,3$ and $t$, $\vec{x}$ are the dimensionful time and position as in (\ref{dim})) and a Lorentz geometry with dimensionless metric 
\begin{equation}
\label{metric3d}
g_{\mu\nu}=\left(
\begin{array}{cccc}
 \frac{4 r^2 f_2^2(r,\lambda)}{\lambda^2} & 0 & 0 & 0 \\
 0 & -1& 0 & 0 \\
 0 & 0 & -1& 0 \\
 0 & 0 & 0 & -1 \\
\end{array}
\right).
\end{equation}
A natural choice for the parameterisation of the geodesics then also presents itself as the proper time 
\begin{equation}
\label{propt}
\frac{d\tau}{dt}=\frac{1}{v_0}\sqrt{g_{\mu\nu}\frac{dx^\mu}{dt} \frac{dx^\nu}{dt}},
\end{equation}
where $\vec{x}(t)$ is a curve solving the equation (\ref{eqmfd}).  Using this, standard vector algebraic identities and the fact that the energy (\ref{constbd}) is constant on this curve, one can rewrite the equation of motion (\ref{eqmfd}) as
\begin{equation}
\frac{d^2\vec{x}}{d\tau^2}=\Omega\left[\vec{x}+\frac{1}{v_0}\left(\frac{d\vec{x}}{d\tau}\times\vec{x}\right)+\frac{1}{v_0^2}\left(\vec{x}\cdot \frac{d\vec{x}}{d\tau}\right)\frac{d\vec{x}}{d\tau}\right]
\end{equation}
and
\begin{equation}
\frac{d^2 x^0}{d\tau^2}=-\frac{v_0}{r\left(g_2(r,\lambda)-\frac{e}{mv_0^2}\right)^2}\frac{dg_2(r,\lambda)}{dr}\vec{x}\cdot \frac{d\vec{x}}{d\tau}
\end{equation}
with
\begin{equation}
\Omega=v_0^2\left[\frac{g_1(r,\lambda)}{\lambda^2}-\frac{1}{r\left(g_2(r,\lambda)-\frac{e}{mv_0^2}\right)}\frac{dg_2(r,\lambda)}{dr}\right].
\end{equation}

This can be interpreted as a geodesic equation 
\begin{equation}
\label{geo}
 \frac{d^2 x^\mu}{d\tau^2}+\Gamma^\mu_{\lambda\nu}\frac{d x^\lambda}{d\tau}\frac{d x^\nu}{d\tau}=0,
 \end{equation}
 with connections in the current coordinate system, which also serves as the fiducial system,  explicitly given by 
 \begin{eqnarray}
 \label{conn}
 \Gamma^i_{00}&=&-\frac{\Omega}{v_0^2}\left(g_2(r,\lambda)-\frac{e}{mv_0^2}\right)^2 x^i,\nonumber\\
  \Gamma^i_{0j}= \Gamma^i_{j0}&=&-\frac{\Omega}{v_0^2}\left(g_2(r,\lambda))-\frac{e}{mv_0^2}\right)\epsilon_{ijk}x^k\nonumber\\
  \Gamma^i_{jk}&=&\frac{\Omega}{2v_0^2}\left( x_j\delta^i_k+x_k\delta^i_j\right)\nonumber,\\
  \Gamma^0_{0j}= \Gamma^0_{j0}&=&-\frac{1}{r\left(g_2(r,\lambda)-\frac{e}{mv_0^2}\right)}\frac{dg_2(r,\lambda)}{dr}x_j.
  \end{eqnarray}
 All other connections vanish.  As usual Latin symbols are reserved for spatial indices and $x_\mu=g_{\mu\nu}x^\nu$, which implies $x_j=-x^j$.  
  
 It is important to note that these connections are not the Levi-Civita connections of the metric (\ref{metric3d}).  Let us also introduce these
  \begin{eqnarray}
  \label{LCconn}
 \tilde{ \Gamma}^0_{0j}&=&-\frac{g_1(r,\lambda)}{\lambda^2}x_j,\nonumber\\
  \tilde{\Gamma}^i_{00}&=&\frac{4r^2f_2^2(r,\lambda)g_1(r,\lambda)}{\lambda^4}x^i,
  \end{eqnarray}
 where we have used  (\ref{eqmfda}). All other connections vanish.  We can now rewrite (\ref{geo}) as 
  \begin{equation}
\label{geo1}
\frac{\tilde{D}}{d\tau}\frac{dx^\mu}{d\tau}= \frac{d^2 x^\mu}{d\tau^2}+\tilde{\Gamma}^\mu_{\lambda\nu}\frac{d x^\lambda}{d\tau}\frac{d x^\nu}{d\tau}=S^\mu_{\lambda\nu}\frac{d x^\lambda}{d\tau}\frac{d x^\nu}{d\tau}\equiv F^\mu,
 \end{equation}
where $S^\mu_{\lambda\nu}=\tilde{\Gamma}^\mu_{\lambda\nu}-\Gamma^\mu_{\lambda\nu}$ and we have introduced the covariant derivative $\frac{\tilde{D}}{d\tau}=\frac{d}{d\tau}+\tilde{\Gamma}$.  Since $S^\mu_{\lambda\nu}$ is written in terms of the dynamical degrees of freedom only, it will transform as a (1,2) tensor under a general coordinate transformation and $F^\mu$ as a 4-vector so that (\ref{geo1}) constitutes a covariant equation \cite{Jackiw}.  The right hand side represents a 4-force that acts on a particle moving in the space-time with a Lorentz geometry (\ref{metric3d}).  The quantities appearing in (\ref{geo1}) are those in the fiducial system and transform in the standard way to any other coordinate system.  Also note that the energy labels the trajectories in the fiducial system.

There is another important consistency check that we must do. From (\ref{geo1}) it follows that
\begin{equation}
\frac{\tilde{D}}{d\tau}\left( \frac{d x^\nu}{d\tau}\frac{d x_\nu}{d\tau}\right)=2\left(\frac{\tilde{D}}{d\tau}\frac{d x^\nu}{d\tau}\right)\frac{d x_\nu}{d\tau}=2F^\nu\frac{d x_\nu}{d\tau}=0.
\end{equation}
An explicit calculation using $F^\mu$ as defined in (\ref{geo1}) shows that this is indeed the case.   

The physically interesting and relevant scenario is one in which (\ref{geo1}) represents geodesic motion ($F^\mu=0$) on some background geometry, possibly perturbed by noncommutative corrections.  These noncommutative perturbations may come from corrections to the metric or through a nonvanishing, noncommutative 4-force or both.  If this is the case, (\ref{geo1}) would represent general relativistic dynamics perturbed by noncommutative corrections, which may open up the possibility of observation.  The scenario we would therefore like to explore is one in which the 4-force vanishes in the commutative limit in (\ref{geo1}).  As already pointed out, the commutative limit amounts to keeping only the leading order terms $\tilde{f}_i(r)$, $\tilde{g}_i(r)$ for the functions $f_i(r,\lambda)$ and $g_i(r,\lambda)$.  For the geometry and 4-force as in (\ref{geo1}) it is then simple to see that the 4-force can only vanish if $\tilde{g}_1(r,\lambda)=0$ and $\frac{d\tilde{g}_2(r,\lambda)}{dr}=0$.   From (\ref{exp}) this requires 
\begin{eqnarray}
\label{geo23d}
&&\tilde{f}_2(r)=\frac{\alpha}{r},\nonumber\\
&&r\left(\tilde{f}_1(r)+\tilde{f}_3(r)\right)+\frac{\tilde{V}(r)}{mv_0^2}=\beta
\end{eqnarray}
for some constants $\alpha,\beta$.   Under these conditions the metric (\ref{metric3d}) reduces in the commutative limit and  after an appropriate rescaling of time to the Minkowski metric and the 4-force in (\ref{geo1}) vanishes.  Of course, the dynamics is rather trivial in this case as all connections also vanish.  This implies that a corresponding choice of the noncommutative Hamiltonian (\ref{ham3}) that leads to (\ref{geo23d}) will give rise to geodesic motion on Minkowski space-time.   In particular this happens for the free particle (\ref{free3d}).    In this case the functions $f_i(r,\lambda)$ can be explicitly computed as in \cite{scholtz1} to give
\begin{eqnarray}
f_1(r,\lambda)&=&\frac{\lambda}{2r}-\frac{\lambda^2}{r^2},\nonumber\\
f_2(r,\lambda)&=&\frac{\lambda}{2r}-\frac{\lambda^2}{4r^2}-\frac{\lambda^3}{16r^3}+O(\lambda^4),\nonumber\\
f_3(r,\lambda)&=&\frac{\lambda}{2r}.
\end{eqnarray} 
Note that there are no higher order corrections to $f_1(r,\lambda)$ and $f_3(r,\lambda)$.   Using this in (\ref{metric3d}) leads precisely to the Minkowski metric with noncommutative corrections.  In addition the 4-force in (\ref{geo1}) also acquires noncommutative corrections.  These corrections vanish in the commutative limit.  

Above we have chosen the most obvious and natural metric with accompanying 4-force, but the possibility is not excluded that other choices of the noncommutative Hamiltonian (functions $f_i(r,\lambda)$), associated with other metrics and 4-forces that still vanish in the commutative limit, can be found.  It is, however, a non-trivial exercise to establish the existence of such Hamiltonians and associated metrics and even more challenging to carry out their explicit construction if they do exist.  The exploration of this possibility will therefore be done elsewhere.  The main point we wish to convey here, explicitly demonstrated in the case of a free particle, is the possibility of a correspondence between dynamics on three dimensional fuzzy space and geodesic motion on some space-time geometry, possibly perturbed by noncommutative corrections and in the presence of a 4-force that is also purely noncommutative in origin.  Both the perturbations to the space-time geometry and the 4-force vanish in the commutative limit to yield geodesic motion on the associated space-time geometry.

We note that the path integral action (\ref{action3d}) involves the auxiliary variables $w$.  It is therefore natural to enquire whether there exists an action, written purely in terms of the dynamical degrees of freedom $x_i$, which also yields the equations of motion (\ref{geo1}).  In the case of three dimensional fuzzy space we were not able to construct such a local action, and it is dubious that it exists.  Also note that even if such an action can be found, the two theories are only equivalent on the classical level, while they may be quite different on the quantum level as the two actions may differ substantially off-shell.  One expects that such a non-local action may be obtained by integrating out the auxiliary variables, which appear quadratically, from (\ref{action3d}).  This procedure was already demonstrated for the Moyal plane in \cite{scholtz1}.  Here, however, matters are complicated by the constraint of physicality in (\ref{con3d}).  As pointed out in \cite{Jackiw}, it may be possible to avoid these issues in lower dimensional theories and it therefore seems worthwhile to pursue the above considerations in a 2+1-dimensional setting, which is the aim of the next section.  If this can be done, it should provide us with a clearer physical picture of noncommutative dynamics.

\section{Classical dynamics on the fuzzy sphere ($S^2_\star$)}
\label{fuzzysphere}

In this section we show that the scenario above also plays out in the case of a less trivial Lorentz geometry and that it is indeed also possible to construct a Lorentz invariant action that yields the covariant equations of motion that derive from the noncommutative dynamics.  For this purpose we consider a particle on the fuzzy sphere.  The quantum mechanical treatment has already been discussed in section \ref{NCQM}. 

To derive the path integral action, we require an overcomplete set of coherent states. We first introduce standard $su(2)$ coherent states defined by
\begin{equation}\label{atomic cs}
	|n,z\rangle=\frac{1}{(1+z\bar{z})^{n/2}}e^{z\hat{X}_-}|j=\frac{n}{2},m=\frac{n}{2}\rangle
\end{equation}
where $\hat{X}_{\pm}=\frac{\hat{x}_1 \pm i \hat{x}_2}{2\lambda}$ and $z=\cot(\theta/2)e^{i\phi}$ is the dimensionless stereographic complex coordinate on the sphere. The identity on $\mathcal{H}_c$ can now be resolved as
\begin{equation}
 \label{idclass}
	\hat{I}_c=\int dzd\bar{z}  \,\mu_n(z,\bar{z} )\ket{n,z}\bra{n,z} ~~~~{\rm with}~~~ \mu_n(z,\bar{z} )=\frac{(n+1)}{\pi(1+z\bar{z} )^2}.
\end{equation}
Correspondingly, we introduce coherent states on ${\cal H}_q$:
\begin{equation}
	|{n,z,w})=\ket{n,z}\bra{n,w}
\end{equation}
and the resolution of the identity on ${\cal H}_q$
\begin{equation}
		\int dzd\bar{z} \,\mu_n(z,\bar{z} )\int dw d\bar{w}\,\mu_n(w,\bar{w})~ |{n,z,w})({n,z,w}|={\bf 1}_q.
\end{equation}

The dimensionless path integral action can now be computed as before with the Hamiltonian (\ref{hams}) to yield
\begin{equation}\label{finalcsaction}
S=\int dT \bigg[~\frac{iR}{2} \bigg[\left(\frac{\dot{z}\bar{z}-\dot{\bar{z}}z}{1 +|z|^2} \right)+\left(\frac{\dot{\bar{w}}w-\dot{w}\bar{w}}{1 +|w|^2}\right) \bigg] + \frac{1}{4} \bigg[\left(\frac{1 -|z|^2}{1 +|z|^2} \right)\left(\frac{1 -|w|^2}{1 +|w|^2}\right) + 2 \frac{\bar{z}w+z\bar{w}}{(1 +|z|^2)(1 +|w|^2)} \bigg]\bigg].
\end{equation}
Here we have defined a dimensionless time as in (\ref{dim}) and also introduced the dimensionless coordinate $R=\frac{r}{\lambda}\approx n$.  As before, the over-dot represents derivation with respect to $T$.  

There are two conserved quantities, the angular momentum and energy.  They are respectively given in dimensionless form by
\begin{eqnarray}
L&=&R\left[\frac{z\bar{z}}{1 +|z|^2} -\frac{w\bar{w}}{1 +|w|^2}\right],\nonumber\\
E&=&\frac{1}{4}-\frac{1}{4} \bigg[\left(\frac{1 -|z|^2}{1 +|z|^2} \right)\left(\frac{1 -|w|^2}{1 +|w|^2}\right) + 2 \frac{\bar{z}w+z\bar{w}}{(1 +|z|^2)(1 +|w|^2)}\bigg],
\end{eqnarray}
where we have been careful to also keep the additive dimensionless constant to the energy in (\ref{hams}).

As before, the equations of motion that interest us are those of the stereographic coordinates $z$ that describes the motion of the particle on the sphere.  To extract these, we must eliminate the $w$ from the equations of motion.  A straightforward computation yields
\begin{equation}\label{eqnz}
\frac{d^2z}{dT^2}= \frac{2\bar{z}}{1+|z|^2}\left(\frac{dz}{dT}\right)^2 -\frac{i}{2R}\frac{dz}{dT}\left(\sqrt{1-\frac{16R^2}{(1+|z|^2)^2}\frac{dz}{dT}\frac{d\bar{z}}{dT}}-1\right)
\end{equation}
and the corresponding complex conjugate for $\bar{z}$. As before there are two branches, but we only focus on the branch with the appropriate commutative limit.

The corresponding dimensionless conserved quantities can also be computed easily
\begin{eqnarray}
\label{constmd}
L&=&\frac{2iR^2}{(1+|z|^2)^2}\left(\bar{z}\frac{dz}{dT}-z\frac{d\bar{z}}{dT}\right)+\frac{R(|z|^2-1)}{2(|z|^2+1)}\left(1-\sqrt{1-\frac{16R^2}{(1+|z|^2)^2}\frac{dz}{dT}\frac{d\bar{z}}{dT}}\right),\nonumber\\
E&=&\frac{1}{4}\left[1-\sqrt{1-\frac{16R^2}{(1+|z|^2)^2}\frac{dz}{dT}\frac{d\bar{z}}{dT}}~\right].
\end{eqnarray}

To highlight the physical content of these equations, it is again convenient to restore dimensions using (\ref{dim}) and to introduce the dimensionful coordinate $\tilde{z}=rz=r\cot(\theta/2)e^{i\phi}$.  For notational convenience, we'll drop the tilde in what follows, but it should be kept in mind that the coordinates are now dimensionful. Doing this, the dimensionful equation of motion reads
\begin{equation}\label{eqnzd}
\frac{d^2z}{dt^2}= \frac{2\bar{z}}{r^2+|z|^2}\left(\frac{dz}{dt}\right)^2 -\frac{iv_0}{2r}\frac{dz}{dt}\left(\sqrt{1-\frac{16r^4}{v_0^2(r^2+|z|^2)^2}\frac{dz}{dt}\frac{d\bar{z}}{dt}}-1 \right),
\end{equation}
where $r$, $t$ and $z$ are the dimensionful radius, time and spatial coordinates respectively and $v_0$ the limiting speed defined in (\ref{speedlimit}).  The corresponding dimensionful angular momentum and energy are given by
\begin{eqnarray}
\ell&=&\frac{2imr^4}{(r^2+|z|^2)^2}\left(\bar{z}\frac{dz}{dt}-z\frac{d\bar{z}}{dt}\right)+\frac{mv_0r(|z|^2-r^2)}{2(|z|^2+r^2)}\left(1-\sqrt{1-\frac{16r^4}{v_0^2(r^2+|z|^2)^2}\frac{dz}{dt}\frac{d\bar{z}}{dt}}\right),\label{constmdda}\\
e&=&\frac{mv_0^2}{4}\left[1-\sqrt{1-\frac{16r^4}{v_0^2(r^2+|z|^2)^2}\frac{dz}{dt}\frac{d\bar{z}}{dt}}\right].\label{constmddb}
\end{eqnarray}
As for $\mathbb{R}^3_\star$ equations (\ref{eqnzd})-(\ref{constmddb}) suggest that the classical ($\hbar\rightarrow 0$) and commutative ($\lambda\rightarrow 0$) limits should be taken such that $\frac{\hbar}{m\lambda}=v_0$ is fixed.  Note, however, that in this case $\lambda$ only features in $v_0$ and this limit is therefore irrelevant as far as equations (\ref{eqnzd})-(\ref{constmddb}) are concerned.   The commutative limit is again obtained by taking the $v_0\rightarrow\infty$ limit ($\lambda\rightarrow 0$ followed by $\hbar\rightarrow 0$), in which case the second term in (\ref{eqnzd}) drops out and the equation of motion of a particle on the commutative sphere results. The corresponding expressions for the angular momentum and energy follow from (\ref{constmdda}) and (\ref{constmddb}).  In particular note that the second term in the expression for the angular momentum drops out in this limit.  Also note that upon expanding the square root in the energy, we obtain the energy to leading order in $\frac{1}{v_0}$ as
\begin{equation}
e=\frac{2mr^4}{(r^2+|z|^2)^2}\frac{dz}{dt}\frac{d\bar{z}}{dt}+O\left(\frac{1}{v_0^2}\right),
\end{equation}
which is precisely the energy of a free particle on the commutative sphere expressed in dimensionful stereographic coordinates  $z=r\cot(\theta/2)e^{i\phi}$.  

We note from (\ref{eqnzd})-(\ref{constmddb}) that the natural speed that appears in these equations is actually $\frac{v_0}{2}$ rather than $v_0$ itself.  For the purposes of writing these equations in a covariant form, it is therefore, in contrast to $\mathbb{R}^3_\star$, more natural to introduce the dimensionful coordinate $x^0=\frac{v_0}{2}t$ and the 3-vector $x^\mu=(x^0,z,\bar{z})$, $\mu=0,z,\bar{z}$. Upon doing this, inspection of these equations again suggest a natural reparameterisation of time as in (\ref{propt}), but now with the metric
\begin{equation}\label{metr}
g_{\mu\nu} = \begin{pmatrix}
1 & 0 & 0  \\ 
0 & 0 & -\frac{2r^4}{(r^2+|z|^2)^2} \\
0 &  -\frac{2r^4}{(r^2+|z|^2)^2} & 0  
\end{pmatrix}
\end{equation}
which represents the space-time metric of a commutative manifold $\mathbb{R} \times S ^2$, where $\mathbb{R}$ corresponds to time and $S^2$ the commutative sphere.  Furthermore we note from (\ref{constmddb}) that
\begin{equation}
\label{propertime}
\frac{dx^0}{d\tau}=\frac{mv_0^3}{2(mv_0^2-4e)}
\end{equation}
where $d\tau=\frac{2}{v_0} \sqrt{g_{\mu\nu}dx^{\mu} dx^{\nu}}$ is the proper time associated with the metric (\ref{metr}).

Using the fact that energy is conserved we have from (\ref{propertime}) and (\ref{eqnzd})
\begin{eqnarray}
&&\frac{d^2x^0}{d\tau^2}=0 \label{eqnt}\\
&&\frac{d^2z}{d\tau^2}- \frac{2\bar{z}}{r^2+|z|^2}\left(\frac{dz}{d\tau}\right)^2 -\frac{4ie}{mv_0^2r}\frac{dz}{d\tau}\frac{dx^0}{d\tau}=0\label{eqm2}.
\end{eqnarray}

It will be advantageous at this stage to use a coordinate system in which the metric takes a diagonal form and this is trivially obtained by splitting (\ref{eqm2}) and its complex conjugate into real and imaginary parts by setting $z=x+iy$. This yields for the metric
\begin{equation}\label{metric}
g_{\mu\nu} = \begin{pmatrix}
1 & 0 & 0  \\ 
0 & -\frac{4r^4}{(r^2+x^2+y^2)^2} & 0 \\    
0 &  0 & -\frac{4r^4}{(r^2+x^2+y^2)^2}  
\end{pmatrix} 
\end{equation}
where $x^{\mu}=(x^0,x,y)$. The pair of equations (\ref{eqm2}) then takes the form
\begin{equation}\label{eqnx}
\frac{d^2x}{d\tau^2}-\frac{2}{r^2+x^2+y^2}\left(x\left(\frac{dx}{d\tau}\right)^2-x\left(\frac{dy}{d\tau}\right)^2+2y\left(\frac{dx}{d\tau}\right)\left(\frac{dy}{d\tau}\right)\right)=-\frac{2ev_0}{r(mv_0^2-4e)}\left(\frac{dy}{d\tau}\right)
\end{equation}
and
\begin{equation}\label{eqny}
\frac{d^2y}{d\tau^2}-\frac{2}{r^2+x^2+y^2}\left(y\left(\frac{dy}{d\tau}\right)^2-y\left(\frac{dx}{d\tau}\right)^2+2x\left(\frac{dx}{d\tau}\right)\left(\frac{dy}{d\tau}\right)\right)=\frac{2ev_0}{r(mv_0^2-4e)}\left(\frac{dx}{d\tau}\right).
\end{equation}

The equations (\ref{eqnt}), (\ref{eqnx}) and (\ref{eqny}) can be combined in the form of a geodesic equation
\begin{equation}\label{geo2}
\frac{d^2x^{\mu}}{d\tau^2}+ \Gamma^{\mu}_{\nu\sigma} \frac{dx^{\nu}}{d\tau} \frac{dx^{\sigma}}{d\tau}=0,
\end{equation}
where
\begin{equation}\label{conn}         
\Gamma^{x}_{0y}=\frac{2e}{mv_0^2r}, \Gamma^{y}_{0x}=-\frac{2e}{mv_0^2r},
\Gamma^{x}_{xx}=\Gamma^{y}_{xy}=-\frac{2x}{r^2+x^2+y^2}~,~\Gamma^{y}_{yy}=\Gamma^{x}_{xy}=-\frac{2y}{r^2+x^2+y^2}
\end{equation}
\begin{equation}
\Gamma^{x}_{yy}=\frac{2x}{r^2+x^2+y^2}, ~\Gamma^{y}_{xx}=\frac{2y}{r^2+x^2+y^2} \nonumber
\end{equation}
and all other $\Gamma^{\mu}_{\nu\sigma}$'s vanish. Note that the spacetime coordinates $x^{\mu}$ can be regarded as the those associated with the fiducial frame where the metric takes the form (\ref{metric}) and (\ref{geo2}) represents the equation of motion of a particle having energy $e$, as measured and observed by an observer in this fiducial frame.   The point to be noted, however, is that, as for $\mathbb{R}^3_\star$, the connection components $\Gamma^{\mu}_{\nu\sigma}$'s appearing in (\ref{geo2}) are not those of the metric (\ref{metric}) or any other metric in the conventional manner. In other words, this connection is not metric-compatible. The components $\tilde{\Gamma}$'s corresponding to the metric (\ref{metric}) can be calculated in the standard way from (\ref{metric}) yielding for the non-zero components 
\begin{equation}
\tilde{\Gamma}^{x}_{xx}=\tilde{\Gamma}^{y}_{xy}=-\frac{2x}{r^2+x^2+y^2}~,~\tilde{\Gamma}^{y}_{yy}=\tilde{\Gamma}^{x}_{xy}=-\frac{2y}{r^2+x^2+y^2}~,~\tilde{\Gamma}^{x}_{yy}=\frac{2x}{r^2+x^2+y^2},\tilde{\Gamma}^{y}_{xx}=\frac{2y}{r^2+x^2+y^2}.
\end{equation}

As for  $\mathbb{R}^3_\star$ we can recast (\ref{geo2}) in the following form
\begin{equation}\label{geodev}
\frac{d^2x^{\mu}}{d\tau^2}+ \tilde{\Gamma}^{\mu}_{\nu\sigma} \frac{dx^{\nu}}{d\tau} \frac{dx^{\sigma}}{d\tau}=S^{\mu}_{\nu\sigma} \frac{dx^{\nu}}{d\tau} \frac{dx^{\sigma}}{d\tau}\equiv F^{\mu}~~;~~~~~S^{\mu}_{\nu\sigma}=\tilde{\Gamma}^{\mu}_{\nu\sigma}-\Gamma^{\mu}_{\nu\sigma}
\end{equation}

The above equation can be interpreted as an equation of motion of a particle with energy $e$, as measured in the fiducial frame, and subjected to a 3-force  $F^{\mu}$ that will cause deviations from geodesic motion.  Also note that this 3-force is purely of noncommutative origin.  The components of the 3-force $F^{\mu}$ can be easily read off and are given by
\begin{equation}\label{contrafor}
F^{0}=0 ~, ~~F^{x} = -\frac{2ev_0}{r(mv_0^2-4e)}\frac{dy}{d\tau}~,~~~
 F^{y} =\frac{2ev_0}{r(mv_0^2-4e)}\frac{dx}{d\tau}.
\end{equation}
One can also easily check that $F^{\mu}u_{\mu}$, with $u^{\mu}$ the 3-velocity, vanishes identically.  Note that the 3-force is not specified externally, but is completely written in terms of the dynamical variables.  Therefore it must transform under a coordinate transformation and, from its defining relation, it must do so as a 3-vector \cite{Jackiw}.

Although the structures of all these equations, particularly (\ref{geodev}), have been obtained in a fiducial frame, the remarks above suggest their covariance under arbitrary diffeomorphism. Finally, observe that in the commutative limit, the equation of motion (\ref{geodev})  reduces to that of a free particle on the surface of a commutative sphere $S^2$.

In this case we therefore conclude that the metric acquires no noncommutative corrections such that it's connections are compatible with the entire set of connection coefficients (\ref{conn}). This is indicative of the fact that the general fuzzy sphere does not admit a metric. This also corroborates the observations made in \cite{KK}, where the spectral distances a la Connes on the fuzzy sphere are found to be highly deformed and gives the commutative geodesic distance only in the $n\longrightarrow \infty$ limit of $su(2)$ representation. Here, in some sense, we are able to identify the occurrence of the 3-force $F^\mu$ in (\ref{geodev}) to be responsible for this. 

The covariant structure of the equations of motion (\ref{conn}) under diffeomorphism, raises the question whether a suitable diffeomorphism invariant action exists from which (\ref{conn}) follows through a variational analysis. We take this up in the next section.

 \section{Action and conserved quantities for a free particle on the fuzzy sphere}
 \label{action}
We begin with the line element corresponding to the metric (\ref{metric}) written as
\begin{equation}
ds^2=\frac{1}{4}v_0^2dt^2+g_{ij}dx^i dx^j; ~~g_{xx}=g_{yy}=-\frac{4r^4}{(r^2+x^2+y^2)^2}.
\end{equation}
Since $z=r\cot(\theta/2)e^{i\phi}$, we can rewrite it in a more familiar form using polar coordinates as
\begin{equation}
ds^2=\frac{v_0^2}{4}dt^2-r^2(d\theta^2+\sin^2{\theta} d\phi^2).
\end{equation}

We introduce the following action, given by the line integral along the world line of a particle with energy $e$ as measured in the fiducial frame:
\begin{equation}\label{actio}
S_e=-\frac{mv_0}{2}\int d\tau \sqrt{g_{\mu\nu}\dot{x}^{\mu} \dot{x}^{\nu}} + \frac{2m}{v_0} \int d\tau a_{\mu}(x)\dot{x}^{\mu}.
\end{equation}
Here $d\tau$ is the proper time associated with the metric (\ref{metric}) and the overdot denotes derivation with respect to $\tau$.  

It is now easy to verify that the equation of motion (\ref{geo2}) results upon setting $\frac{\delta S}{\delta x^{\lambda}}=0$:
\begin{equation}\label{twoform}
\frac{d}{d\tau}(\frac{\dot{x}^\mu}{N}) + \frac{1}{N}\tilde{\Gamma}^{\mu}_{\nu\sigma} \dot{x}^{\nu}\dot{x}^{\lambda}=\frac{4}{v_0^2}{F^{\mu}}_{\nu}\dot{x}^{\nu} ~;~~~~~F_{\lambda\mu}=\partial_{\lambda}a_{\mu}-\partial_{\mu}a_{\lambda}.
\end{equation}
Here $N=\sqrt{g_{\mu\nu}\dot{x}^{\mu} \dot{x}^{\nu}}=\frac{v_0}{2}$.~~This equation can be rewritten more compactly in terms of the covariant derivative $\frac{D}{D\tau}=\frac{d}{d\tau}+\tilde{\Gamma}$ as
\begin{equation}\label{maineq}
\frac{D}{D\tau}(\dot{x}^\mu)= F^{\mu}~~~~~~~~~~;~~F^{\mu}\equiv\frac{2}{v_0}{F^{\mu}}_{\nu}\dot{x}^{\nu} .
\end{equation}
The above field two-form components $F_{\sigma\mu}$ can be written in the fiducial frame as
\begin{equation}
F_{\sigma\mu}=Q_e\sqrt{g}\epsilon_{\sigma\mu\lambda}\eta^{\lambda}~~~~~;~~\eta^{\lambda}=(1,0,0)
\end{equation}
where $\eta^{\lambda}$ is a unit time-like vector representing, up to a constant, the 3-velocity vector of a particle at rest. The corresponding covariant components of
the 3-force are obtained as $F_{\sigma}= \frac{2}{v_0}F_{\sigma\mu}\dot{x}^{\mu} = \frac{2}{v_0}Q_e \sqrt{g}\epsilon_{\sigma\mu\lambda}\eta^{\lambda}\dot{x}^{\mu}$. Here $Q_e$ is an effective charge and can be identified by using (\ref{eqnx},\ref{eqny}) as
\begin{equation}\label{charge}
 Q_e=\frac{ev_0^2}{r(mv_0^2-4e)}.
\end{equation}
Finally, and remarkably, on expressing this field 2-form in polar coordinates $z=r\cot(\theta/2)e^{i\phi}$ we find it to be given by
\begin{equation}\label{U1}
F(x)=\frac{1}{2} F_{\mu\nu}(x)dx^{\mu}\wedge dx^{\nu}=r^2Q_e \sin{\theta}d\theta \wedge d\phi,
\end{equation}
which we readily recognise to be the two form of the Dirac magnetic monopole. Consequently, the connection 1-form $a(x)\equiv a_\mu(x)dx^\mu$ satisfying $F=da$ in (\ref{twoform}) can be written, with suitable choice of gauge, as 
\begin{equation}\label{U0}
a(x)=r^2Q_e(\pm 1 - \cos{\theta})d\phi
\end{equation}
for, respectively, the northern/southern hemisphere of the commutative sphere $S^2$ \cite{yang}. This indicates that the on-shell dynamics of a free particle on the fuzzy sphere $S^2_{\star}$ is equivalent to the on-shell dynamics of a particle with charge $Q_e$ in (\ref{charge}), moving on the commutative sphere and coupled to the background monopole field. Note that the subscript $e$ in the action $S_e$ (\ref{actio}) and the effective charge $Q_e$ serves just as a label and will not change under diffeomorphism transformation, as it refers to the energy measured in the fiducial frame only.

Greater insight into the relation between (\ref{actio}) and (\ref{finalcsaction}) can be obtained by considering the relation between the conserved quantities following from them.  In the case of (\ref{finalcsaction}) these have already been computed and are given by (\ref{constmdda}) and (\ref{constmddb}).  We therefore now proceed to compute the conserved quantities for (\ref{actio}).  There are two ways we can go about this, the first is to compute the conserved quantities using the Noether construction and the second is to compute the energy momentum tensor and from there the conserved quantities.  However, as was pointed out in \cite{Jackiw}, the latter is difficult to implement as the connection one-form $a(x)$ (\ref{U0}) has only an implicit dependence on the metric $g_{\mu\nu}$ arising from the fact that the field 2-form $F$ is proportional to the area form of commutative $S^2$. We therefore follow Noether's prescription to obtain these conserved quantities, but for completeness we also construct the complete energy-momentum tensor in the sequel using the method described in \cite{Jackiw} and compare with the results of the Noether prescription.

Before proceeding a number of observations are necessary.  Note that since (\ref{actio}) and (\ref{finalcsaction}) have the same on-shell dynamics, a quantity that is conserved on-shell (the equations of motion are used to show that its time derivative vanishes) in the one, will also be conserved on-shell in the other.   Despite this, there is no a priori reason why the conserved quantities computed from these two actions will coincide as these actions differ off-shell and also typically by total time derivatives.  In the light of our previous remark, one would then expect that the conserved quantities derived from (\ref{actio}) will generally be functions of those derived from (\ref{finalcsaction}).  This is hardly surprising as once a conserved quantity has been obtained, any function of this conserved quantity is also conserved, but the functional form may depend on the action from which it has been derived.

Let us first compute the energy $\mathcal{E}$ from the action (\ref{actio}) by following Noether's prescription. This is tantamount to carrying out a Legendre transformation of the Lagrangian $L_e$
\begin{equation}
\mathcal{E}=\dot{x}^i p_i-L_e~~~;~ i=1,2
\end{equation}
where,
\begin{eqnarray}
L_e&=&-\frac{mv_0}{2}\sqrt{g_{\mu\nu}\dot{x}^{\mu} \dot{x}^{\nu}} + \frac{2m}{v_0} a_{\mu}(x)\dot{x}^{\mu}\nonumber\\
&=&-\frac{mv_0}{2} \sqrt{\frac{v_0^2}{4}-r^2(\dot{\theta}^2+\sin^2{\theta} \dot{\phi}^2)} - \frac{2m}{v_0} r^2Q_e\cos{\theta} \dot{\phi} \label{lagrangianpolar}
\end{eqnarray}
and we have omitted a total derivative term when writing the Lagrangian in polar coordinates. The energy comes out to be just the relativistic energy of a free particle on the commutative sphere
\begin{equation}\label{fullrelenergy}
\mathcal{E}=\frac{mv_0^2}{4\sqrt{1+4g_{xx}\frac{\dot{\vec{x}}^2}{v_0^2}}}.
\end{equation}
Note that the interaction term i.e. the second term in $L_e$ does not contribute to $\mathcal{E}$, as it is linear in velocity.   

For calculating the conserved angular momentum from the action using Noether's prescription, it will be easier to work with the Lagrangian (\ref{lagrangianpolar}) written in polar coordinates. We note that $\phi$ is cyclic and so will lead to conservation of the corresponding conjugate momentum $J$ which is nothing but the angular momentum given by
\begin{equation}
\label{fullangmom}
J=-\frac{m\sqrt{g}\epsilon_{ij}x^i\dot{x}^j}{\sqrt{1+4g_{xx}\frac{\dot{\vec{x}}^2}{v_0^2}}}+\frac{2mev_0r}{mv_0^2-4e}\cos\theta.
\end{equation}

Comparing (\ref{fullrelenergy}), (\ref{fullangmom}) with (\ref{constmdda}) and (\ref{constmddb}), we find the following relation between the constants of motion of the two systems
\begin{eqnarray}
\label{constmrel}
\mathcal{E}&=&\frac{mv_0^2}{4\left(1-\frac{4e}{mv_0^2}\right)},\nonumber\\
J&=&\left(\frac{mv_0^2}{mv_0^2-4e}\right)\ell.
\end{eqnarray}
We note that in the $v_0\rightarrow\infty$ limit these quantities coincide.  Interestingly, this suggests that the commutative limit of (\ref{finalcsaction}) coincides with the nonrelativistic limit of (\ref{actio}), and both are then the action of a free particle on the sphere.  

Finally, as a benchmark, we would like to obtain the covariant energy-momentum (EM) tensor and evaluate it in the fiducial frame. As mentioned above, $T^{\mu \nu}$ will have contributions from both the free point particle action, i.e. the first term in (\ref{actio}), and also from the monopole term of the action since the associated force here is a geometrical gravitational force in the sense of \cite{Jackiw}.  As explained in  \cite{Jackiw} and also mentioned above, the connection $a_\mu$ has only an implicit dependence on the metric and no explicit dependence. This complicates the calculation, but we can follow the same strategy as in \cite{Jackiw} where the (1+1) dimensional case was considered.  We take as an ansatz for the energy-momentum tensor $T^{\mu \nu}$
\begin{equation}\label{em0}
T^{\mu \nu}=\frac{mv_0^2}{4}\int d\tau \frac{\dot{x}^{\mu}(\tau)\dot{x}^{\nu}(\tau)}{\sqrt{g}N}\delta^3(x-x(\tau))+\Lambda(x) g^{\mu\nu}+ \xi(x) G^{\mu\nu}
\end{equation}
where $G^{\mu\nu}\equiv(R^{\mu\nu}-\frac{1}{2}g^{\mu\nu}R)$ is the Einstein tensor and the only other symmetric tensor, apart from the metric, which is covariantly conserved.  The arbitrary functions $\Lambda(x)$ and $\xi(x)$ are still to be determined.  Since the EM tensor must be covariantly conserved, we have
\begin{equation}
\grad{\mu}T^{\mu \nu}=\frac{mv_0^2}{4}\int d\tau \frac{\dot{x}^{\mu}(\tau)\dot{x}^{\nu}(\tau)}{\sqrt{g}N}\grad_{\mu}\delta^3(x-x(\tau))+\partial_{\mu}\Lambda(x) g^{\mu\nu}+ \partial_{\mu}\xi(x) G^{\mu\nu}=0.
\end{equation}
This implies, upon integrating by parts, and using the on-shell equations of motion (\ref{maineq})
\begin{equation}\label{emi}
m\int d\tau~ Q_e\epsilon_{\alpha\rho\lambda}\eta^{\lambda}\dot{x}^{\rho}(\tau)\delta^3(x-x(\tau))+\partial_{\alpha}\Lambda(x) + \partial_{\mu}\xi(x) {G^{\mu}}_{\alpha}=0.
\end{equation}

The components of the Einstein tensor ${G^{\mu}}_{\nu}$ for the metric (\ref{metric}) are
\begin{equation}
{G^{\mu}}_{\nu}= \begin{pmatrix}
\frac{-R}{2} & 0 & 0  \\ 
0 & 0 & 0 \\
0 &  0 & 0  \
\end{pmatrix}
\end{equation}
where $R=g^{\mu\nu}R_{\mu\nu}=-\frac{2}{r^2}$ is the scalar curvature of the manifold $\mathbb{R} \times S ^2$. Setting $\alpha=0,\,i\;  (i=1,2)$ in (\ref{emi}), readily yields for the temporal and spatial components
\begin{equation}\label{con1}
\partial_{0}\Lambda(x) - \frac{R}{2} \partial_{0}\xi(x)=0,
\end{equation}

\begin{equation}\label{con2}
\partial_{i}\Lambda(x) = -m\int d\tau  Q_e\epsilon_{ij}\dot{x}^{j}(\tau) \delta^3(x-x(\tau)).
\end{equation}

Integrating (\ref{con1}), we get
\begin{equation}
\xi(x)=\frac{2}{R}( \Lambda(x) - f(\vec{x})),
\end{equation}
where $f(\vec{x})$ is a time-independent function to be determined below.   Similarly, by integrating (\ref{con2}) we get
\begin{equation}
\Lambda(x)=-\frac{2mQ_e}{v_0}\bigg(\dot{x}^2(t) \delta(x^2-x^2(t)) \theta(x^1-x^1(t)) - \dot{x}^1(t)\delta(x^1-x^1(t)) \theta(x^2-x^2(t))\bigg),
\end{equation}
where we have ignored an additive time-dependent constant of integration,
so that (\ref{em0}) now becomes
\begin{equation}\label{em1}
T^{\mu \nu}=\frac{mv_0^2}{4}\int d\tau \frac{\dot{x}^{\mu}(\tau)\dot{x}^{\nu}(\tau)}{\sqrt{g}N}\delta^3(x-x(\tau))+\Lambda(x) \bigg(g^{\mu\nu}+\frac{2}{R}G^{\mu\nu}\bigg) - \frac{2}{R}f(\vec{x}) G^{\mu\nu}.
\end{equation}

Finally, we determine the function $f(\vec{x})$ by identifying the energy $\mathcal{E}$ in (\ref{fullrelenergy}) as 
\begin{equation}
\mathcal{E}=\int d^2x \sqrt{g}~ T^{00}.
\end{equation}
 This readily yields $f(\vec{x})=0$.
The complete expression of the EM tensor (\ref{em1}) can now be written down as
\begin{eqnarray}\label{emf}
T^{\mu \nu}=&&\frac{mv_0^2}{4}\int d\tau \frac{\dot{x}^{\mu}(\tau)\dot{x}^{\nu}(\tau)}{\sqrt{g}N}\delta^3(x-x(\tau))\nonumber \\&
&-\frac{2mQ_e}{v_0}(g^{\mu\nu}+\frac{2}{R}G^{\mu\nu})\bigg(\dot{x}^2(t) \delta(x^2-x^2(t)) \theta(x^1-x^1(t)) - \dot{x}^1(t)\delta(x^1-x^1(t)) \theta(x^2-x^2(t))\bigg).
\end{eqnarray}
Now observe that the 00-component of $T^{\mu\nu}$ i.e. $T^{00}$ doesn't receive any contribution from the second term in view of the fact that $g^{00}+\frac{2}{R}G^{00}=0$. Consequently we can identify $\mathcal{E}=\int d^2x \sqrt{g}~ T^{00}.$ This is, however, not true anymore for the angular momentum
\begin{equation}
\tilde{J}\equiv \int d^2x \sqrt{g}\bigg(x^1 T^{02}-x^2 T^{01}\bigg)\neq J.
\end{equation}
This discrepancy should not be surprising as it is well-known that these two prescriptions of obtaining the EM tensors i.e. Noether's and $T^{\mu\nu}\sim \frac{\delta S}{\delta g_{\mu\nu}}$ are not always equivalent. Sometimes equivalence can be established by adding a suitable total divergence term \emph{a la} Belinfante, but in some cases the difference in the angular momentum expressions, in particular in some (1+2) dimensional systems with a Chern-Simons term, can be interpreted as fractional spin (see for example \cite{annofphys} and references therein).

This result simply emphasises what was already apparent on the level of the equations of motion, namely that these two systems can only be equivalent on-shell in the presence of a coupling to a background magnetic monopole field, which makes an indispensible contribution to the EM tensor. This coupling is the origin of the 3-force in the equation of motion (\ref{geo2}) and purely noncommutative in origin.  The manifest Lorentz covariance of these equations of motion can now also be understood as it arises in the same way as the Lorentz force in normal electromagnetism, where it is well known that it can be cast in a covariant form as above.  As in the case of $\mathbb{R}^3_\star$, the main message here is that the classical dynamics on $S^2_\star$ corresponds to perturbed geodesic motion on a standard Lorentz space-time, where the perturbation is a purely noncommutative 3-force.  

Despite the fact that we were able to derive an action that yields the same equations of motion as the noncommutative action, one must realise that this only implies that the two theories are equivalent on-shell and their constants of motion are generally related through some functional relation as in (\ref{constmrel}). Quantisation of the two theories will generally result in non-equivalent theories as the off-shell behaviour of the actions are quite different.  The equivalence found above between the noncommutative dynamics on the fuzzy sphere and commutative dynamics on the commutative sphere in the presence of a magnetic monopole is therefore restricted to the classical dynamics only.  This does, however, open up the interesting possibility of a non-equivalent quantisation of the commutative dynamics in terms of noncommutative dynamics.    

\section{Conclusions and outlook}
\label{Conclusions}
We have shown, quite remarkably, that the classical equations of motion that follow from the noncommutative path integral action on three dimensional fuzzy space ($\mathbb{R}^3_\star$ ) and the fuzzy sphere ($S^2_\star$) are underpinned by a natural Lorentz geometry. In the case of a free particle (\ref{free3d}) on $\mathbb{R}^3_\star$ this simply turned out to be 4-dimensional Minkowski space-time and in the case of a free particle (\ref{freesp}) on $S^2_\star$, it is given by $R\times S^2$ with metric as in (\ref{metric}).  In both cases the dynamics correspond to geodesic motion perturbed  by a force that is purely noncommutative in origin.  In the case of $\mathbb{R}^3_\star$ the metric also acquired noncommutative corrections, but this is not the case for ($S^2_\star$) where the force is the only manifestation of noncommutativity.  

For $S^2_\star$, we were able to find an action, written only in terms of the coordinates and their time derivatives, that also yields the noncommutative equations of motion.  This action corresponds to a particle moving on the commutative sphere, appropriately coupled to the background field of a magnetic monopole. We also showed that the constants of motion are not identical, but related through a functional relation.  In the $v_0\rightarrow\infty$ they do coincide, which suggests that the commutative limit of the free particle action on $S^2_\star$ coincides with the nonrelativistic limit of this action, both being the action of a free particle on the sphere.    It must, however, be noted that the equivalence between the noncommutative and commutative theories only apply at the classical level and that they may be inequivalent as quantum theories.  This creates the interesting possibility of non-equivalent quantisations of these theories.

These results also open up an interesting relation between classical dynamics on noncommutative spaces and classical theories of gravity.   Here we have derived this correspondence in two cases starting from the noncommutative path integral action.  It will be particularly interesting to find examples where the correspondence can be derived by starting on the gravity side.  In this context it is highly desirable to find the noncommutative spaces and dynamics, if they exist, for which the classical equations of motion are underpinned by Lorentz geometries corresponding to vacuum solutions of the Einstein equations.  This and related issues will be explored elsewhere. 

\section{Acknowledgements}
FGS acknowledges generous support from the S.N. Bose National Centre for Basic Sciences, India where part of this work was completed. One of the authors, S.K.P., would like to thank UGC-India for providing financial support in the form of fellowship during
the course of this work.

%%%%%%%%%%%%%%%%%%%%%%%%%%%%%%%%%%%%%%%%%%%%%%%%%%%%%%%%%%%%%%%%%%%%%%%%%%%%
%%%%%%%%%%%%%%%%%%%%%%%%%%%%%%%%%%%%%%%%%%%%%%%%%%%%%%%%%%%%%%%%%%%%%%%%%%%%
\end{document}